# ANALYSIS OF LOSS NETWORKS WITH ROUTING

By Nelson Antunes, Christine Fricker, Philippe Robert
and Danielle Tibi

*Universidade do Algarve, INRIA, INRIA and Université Paris 7*

This paper analyzes stochastic networks consisting of finite capacity nodes with different classes of requests which move according to some routing policy. The Markov processes describing these networks do not, in general, have reversibility properties, so the explicit expression of their invariant distribution is not known. Kelly's limiting regime is considered: the arrival rates of calls as well as the capacities of the nodes are proportional to a factor going to infinity. It is proved that, in limit, the associated rescaled Markov process converges to a deterministic dynamical system with a unique equilibrium point characterized by a nonstandard fixed point equation.

**1. Introduction.** In this paper, a new class of stochastic networks is introduced and analyzed. Their dynamics combine the key characteristics of the two main classes of queueing networks: loss networks and Jackson type networks.

1. Each node of the network has finite capacity, so a request entering a saturated node is rejected, as in a loss network.
2. Requests visit a subset of nodes along some (possibly) random route, as in Jackson or Kelly's networks.

This class of networks is motivated by the mathematical representation of cellular wireless networks. Such a network is a group of base stations covering some geographical area. The area where *mobile users* communicate with a *base station* is referred to as a *cell*. See Figure 1. A base station is responsible for the bandwidth management concerning mobiles in its cell. New calls are initiated in cells and calls are handed over (transferred) to the corresponding neighboring cell when mobiles move through the network. A new or a handoff

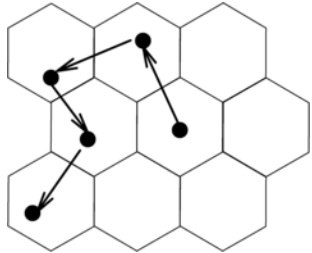

Fig. 1. *The motion of a mobile among the cells of the network.*

call is accepted if there is available bandwidth in the cell, otherwise it is rejected.

Previously, these networks have been modeled at a macroscopic level as loss networks characterized by call arrival rates, mean call lengths, handoff rates and capacity restrictions on the number of calls, in the case of exponential times. One of the main quantities of interest is the *stationary blocking probability* of the network at each node, defined as the stationary probability that a call arriving at that node cannot be accepted. Approximations have been used to analyze these networks; see [1, 5, 11] and the references therein.

Assuming Poisson arrivals and exponentially distributed random variables, the evolution of such a network with $N$ nodes can be represented as a Markov jump process $(X(t))$ with values in some finite (but large) set $\mathcal{S}$. It turns out that, contrary to uncontrolled loss networks with fixed routing, the Markov process $(X(t))$ is not in general reversible or quasi-reversible. Consequently, contrary to Jackson networks and the like, or uncontrolled loss networks, *these networks do not have a stationary distribution with a product form*.

In this paper, the time evolution of these networks is analyzed by considering Kelly's scaling. The arrival rates and capacities at nodes are proportional to some factor $N$ which becomes large. This scaling has been introduced by Kelly [8] to study the invariant distribution of loss networks. A study of the time evolution of loss networks under this scaling has been carried out by Hunt and Kurtz [6]. See [7] for a survey of these questions. A different scaling is considered in [2].

*The equilibrium points.* The time evolution of the network can be (roughly) described as follows. A stochastic process $(\overline{X}_N(t))$ associated with the state of the network for the parameter $N$ is introduced: $\overline{X}_N(t)$ is the vector describing the numbers of customers of different classes at the nodes of the network. The equation of evolution for the network is

$$\frac{d}{dt}\overline{X}_N(t) = F_N(\overline{X}_N(t)) + \overline{M}_N(t), \qquad t \geq 0,$$

where $(\overline{M}_N(t))$ is a martingale which vanishes as $N$ becomes large, $F_N$ is a somewhat complicated functional (associated with the generator of the corresponding Markov process) converging to some limit $F$. As $N$ goes to infinity, it is proved that $(\overline{X}_N(t))$ converges to some function $(x(t))$ satisfying the deterministic equation

$$\frac{d}{dt}x(t) = F(x(t)), \qquad t \geq 0. \tag{1}$$

The equilibrium points of the limiting process are the solutions $x$ of the equation $F(x) = 0$. It is shown in this paper (and this is a difficult point) that there is only one equilibrium point in Kelly's limiting regime.

*Related work.* For classical uncontrolled loss networks, the invariant probability has a product form representation. Nevertheless, the evolution of these networks under Kelly's scaling turns out to be quite intricate. Hunt and Kurtz [6] showed that at any $x$, the vector field $F(x)$ driving the limiting dynamical system is related to some reflected random walk in $\mathbb{R}_+^d$ with jump rates depending on $x$. Intuitively, the situation can be described as follows. At points $x$ at which this random walk is ergodic, $F(x)$ is expressed in terms of its invariant distribution; at $x$ at which the random walk is transient, the exit paths to infinity determine $F(x)$. It is not known, in general, whether there always exists a unique limiting dynamical system. Hunt and Kurtz [6], Bean, Gibbens and Zachary [3, 4] and Zachary [12] analyzed several examples with one or two nodes where uniqueness is shown to hold.

*Results of the paper.* Using the terminology of cellular networks, users arriving in the network correspond to new requests for a connection in a cell. Different classes of customers access the network—classes differ by their arrival rate, by their *dwell time* at the nodes (i.e., the amount of time that a mobile of an ongoing call remains in a given cell), by their call duration and also by their routing through the network. During a call, a user moves from one cell to another according to some Markovian mechanism, depending on his class. When a user moves to another cell (node), this cell has to be nonsaturated to accommodate the user, otherwise the user is rejected (the call is lost). If it is not rejected during the travel through the network, the user call terminates after the call duration time has elapsed.

For the networks analyzed in this paper, the uniqueness of the limiting dynamical system is not difficult to establish. The main difficulty lies in the complexity of the system of equations defining the equilibrium points of the dynamical system. Since there does not seem to exist some reasonably simple contracting scheme to solve these equations, the uniqueness of the equilibrium points is therefore a quite challenging problem. For example, in Section 4.2, the case of a very simple network with two nodes and two deterministic routes is investigated and the explicit representation of the equilibrium point is obtained, expressed in terms of quite complicated polynomial expressions involving the parameters.

The paper is organized as follows. Section 2 introduces the Markovian description of these networks, Section 3 gives the convergence results, together with the description of the limiting dynamical system. Section 4 is devoted to the main results of the paper—it is shown that, in limit, there exists a unique stable point for the network. The ingredients used to obtain this uniqueness result are:

- a dual approach to the problem of uniqueness, that is, finding the set of parameters such that a given point is an equilibrium point of the dynamical system;
- a key inequality proved in the Appendix;
- a convenient probabilistic representation of a set of linear equations.

The inequality proved in the Appendix involves a quantity related to relative entropy, but, curiously, it does not seem to be a consequence of a standard convex inequality as is usually the case in this type of situation.

**2. The stochastic model.** The network consists of a finite set $I$ of nodes, node $i \in I$ having capacity $\lfloor c_i N \rfloor$, where $c_i > 0$ and $N \in \mathbb{N}$. This network receives a finite number of classes of customers, indexed by a finite set $R$; class $r \in R$ customers enter the network according to a Poisson process with rate $\lambda_r N$, where $\lambda_r > 0$.

- *Call duration.* A class $r$ customer who thus far has neither been rejected nor routed to the outside (see *Routing* below) leaves the network after an exponentially distributed time with rate $\mu_r$ (call duration in the context of a cellular network). The case $\mu_r = 0$ is not excluded; it corresponds to the situation where customers stay in the network as long as they are not rejected or routed to the outside.

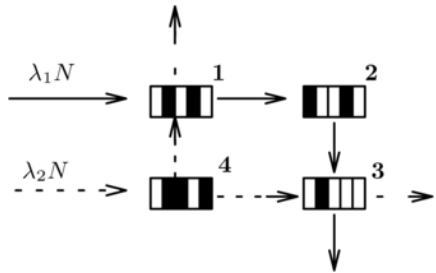

Fig. 2. *A network with two classes of customers.*

- *Dwell time.* The residence time of a customer of class $r$ at any node $i \in I$ is exponentially distributed with parameter $\gamma_r$. Such a customer can leave the node before the end of his dwell time (due to the end of call) at rate $\mu_r$.
- *Routing.* A class $r$ customer entering the network arrives at some random node in $I$ whose distribution is $q_r$ and then moves from one node to another, or to the outside (referred to as node 0), according to some transition matrix $p^{(r)}$ on $I \times I \cup \{0\}$. By changing the parameter of the residence time, it can be assumed without loss of generality that the matrix $p^{(r)}$ is 0 on the diagonal.
- *Capacity requirements.* All customers require one unit of capacity at each node.

All random variables used for arrivals, residence times and dwell times are assumed to be independent.

This class of networks includes the case of classes of customers with deterministic routing, as in Kelly's networks, and also classes of customers with Markovian routing, as in Jackson networks. Figure 2 represents a network with two classes of customers—class 1 customers follow a deterministic route, while class 2 customers can reach either node 1 or node 3 from node 4, and the capacities of the nodes are 5. Note that for the general model, no assumptions have been made concerning the transition matrices $p^{(r)}(\cdot,\cdot)$, so some classes of customers may achieve infinite loops in the network.

*Notation.* For $i \in I$, $r \in R$ and $t \geq 0$, $X_{i,r}^N(t)$ denotes the number of class $r$ customers at node $i$ at time $t$. $(X^N(t)) = (X_{i,r}^N(t), i \in I, r \in R)$ is the corresponding process. The renormalized process is defined as

$$\overline{X}_{i,r}^N(t) = \frac{1}{N} X_{i,r}^N(t)$$

and $\overline{X}^N(t) = (\overline{X}_{i,r}^N(t), i \in I, r \in R)$.

Denote by $I_r \subset I$ the set of nodes which can be visited by a class $r$ customer, that is, $i \in I_r$ when $i$ is visited with positive probability by the Markov chain with transition matrix $p^{(r)}$ and initial distribution $q_r$. It is assumed that $I = \bigcup_{r \in R} I_r$. The state space of the Markov process $(X^N(t))$ is

$$\mathcal{S}_N = \left\{ x = (x_{i,r}) \in \mathbb{N}^{I \times R} : \sum_r x_{i,r} \leq c_i N \text{ and } x_{i,r} = 0 \text{ if } i \notin I_r \right\}.$$

The $Q$-matrix $(A_N(x,y))$ of $(X^N(t))$ is given follows:
*External arrival of a class $r$ customer at node $i$:*

$$A_N(x, x + e_{i,r}) = \lambda_r N q_r(i) \mathbb{1}_{\{x + e_{i,r} \in \mathcal{S}_N\}};$$

*Service completion, rejection by a cell or a transition to the outside*:

$$A_N(x, x - e_{i,r}) = x_{i,r}\left(\mu_r + \gamma_r \sum_{j \in I} p^{(r)}(i,j)\mathbb{1}_{\{x+e_{j,r} \notin \mathcal{S}_N\}} + \gamma_r p^{(r)}(i,0)\right);$$

*Transfer from node $i$ to node $j$*:

$$A_N(x, x - e_{i,r} + e_{j,r}) = \gamma_r x_{i,r} p^{(r)}(i,j)\mathbb{1}_{\{x+e_{j,r} \in \mathcal{S}_N\}},$$

where $e_{i,r}$ is the unit vector at coordinate $(i,r)$. The state space of the renormalized process $(\overline{X}^N(t))$ is given by

$$\mathcal{X}_c = \left\{x = (x_{i,r}) \in \mathbb{R}_+^{I \times R} : \sum_r x_{i,r} \leq c_i \text{ and } x_{i,r} = 0 \text{ if } i \notin I_r\right\},$$

the subscript $c = (c_i)$ of $\mathcal{X}_c$ standing for the vector of capacities.

**3. Convergence results.** The following proposition establishes the deterministic behavior of $\overline{X}^N(t)$ as $N$ goes to infinity. This result is the consequence of the fact that the stochastic perturbations of the original system are of order $\sqrt{N}$ and therefore vanish because of the scaling in $1/N$.

To describe the time evolution of the network, one introduces the following Poisson processes: $\mathcal{N}_\xi$ denotes a Poisson process with parameter $\xi > 0$, and an upper index $\mathcal{N}_\xi^p$, $p \in \mathbb{N}^d$, $d \in \mathbb{N}$, is added when several such Poisson processes are required. For example, for $i \in I$ and $r \in R$, $\mathcal{N}_{\lambda_r q_r(i)}$ is the external arrival Poisson process of class $r$ customers at node $i$. In a similar way, for $k \geq 1$, $\mathcal{N}_{\gamma_r p^{(r)}(i,j)}^k$ is the Poisson process associated with the transfer of the $k$th class $r$ customers from node $i$ to $j \in I \cup \{0\}$.

For $t \geq 0$ and $(i,r) \in I \times R$, let $Y_i^N(t) = \lfloor c_i N \rfloor - \sum_r X_{i,r}^N(t)$. The quantity $Y_i^N(t)$ is the size of the free space at node $i$. The process $(X^N(t))$ can then be represented as the solution of the following stochastic integral equation:

$$
\begin{aligned}
X_{i,r}^N(t) = X_{i,r}^N(0) &+ \int_0^t \mathbb{1}_{\{Y_i^N(s-)>0\}} \mathcal{N}_{\lambda_r N q_r(i)}(ds) \\
&+ \sum_{j \in I - \{i\}} \sum_{k \geq 1} \int_0^t \mathbb{1}_{\{k \leq X_{j,r}^N(s-), Y_i^N(s-)>0\}} \mathcal{N}_{\gamma_r p^{(r)}(j,i)}^k(ds) \\
&- \sum_{\substack{j \in I \cup \{0\} \\ j \neq i}} \sum_{k \geq 1} \int_0^t \mathbb{1}_{\{k \leq X_{i,r}^N(s-)\}} \mathcal{N}_{\gamma_r p^{(r)}(i,j)}^k(ds) \\
&- \sum_{k \geq 1} \int_0^t \mathbb{1}_{\{k \leq X_{i,r}^N(s-)\}} \mathcal{N}_{\mu_r}^{i,k}(ds).
\end{aligned}
$$

(2)

Here $f(t-)$ denotes the limit on the left of the function $f$ at $t$. By compensating the Poisson processes, that is, by replacing the differential term $\mathcal{N}_\xi(ds)$ by the martingale increment $\mathcal{N}_\xi(ds) - \xi\,ds$, one gets the identity

$$
\begin{aligned}
X_{i,r}^N(t) = {} & X_{i,r}^N(0) + M_{i,r}^N(t) + \lambda_r N q_r(i) \int_0^t \mathbb{1}_{\{Y_i^N(s-)>0\}}\,ds \\
& + \gamma_r \sum_{j\in I} p^{(r)}(j,i) \int_0^t X_{j,r}^N(s-)\mathbb{1}_{\{Y_i^N(s-)>0\}}\,ds \\
& - (\gamma_r + \mu_r) \int_0^t X_{i,r}^N(s-)\,ds,
\end{aligned}
\tag{3}
$$

where $(M_{i,r}^N(t))$ is the martingale obtained from the compensated integrals of the previous expression.

Denoting the renormalized martingale $\overline{M}_{i,r}^N(t) = M_{i,r}^N(t)/N$, one finally gets

$$
\begin{aligned}
\overline{X}_{i,r}^N(t) = {} & \overline{X}_{i,r}^N(0) + \overline{M}_{i,r}^N(t) + \lambda_r q_r(i) \int_0^t \mathbb{1}_{\{Y_i^N(s-)>0\}}\,ds \\
& + \gamma_r \sum_{j\in I} p^{(r)}(j,i) \int_0^t \overline{X}_{j,r}^N(s-)\mathbb{1}_{\{Y_i^N(s-)>0\}}\,ds \\
& - (\gamma_r + \mu_r) \int_0^t \overline{X}_{i,r}^N(s-)\,ds.
\end{aligned}
\tag{4}
$$

The evolution equations for the renormalized process having now been written, one can establish the main convergence result:

THEOREM 1. *If the initial state $\overline{X}^N(0)$ converges to $x \in \mathcal{X}_c$ as $N$ goes to infinity, then $(\overline{X}^N(t))$ converges in the Skorohod topology to the solution $(x(t))$ of the following differential equation: For $(i,r) \in I \times R$,*

$$
\frac{d}{dt} x_{i,r}(t) = \left( \lambda_r q_r(i) + \gamma_r \sum_j x_{j,r}(t) p^{(r)}(j,i) \right) \tau_i(x(t)) - (\gamma_r + \mu_r) x_{i,r}(t)
\tag{5}
$$

*with $x(0) = x$ and*

$$
\tau_i(x) = \begin{cases} 1, & \text{if } \sum_r x_{i,r} < c_i, \\ \rho_x^i \wedge 1, & \text{otherwise,} \end{cases}
$$

*where $a \wedge b = \min(a,b)$ for $a, b \in \mathbb{R}$ and*

$$
\rho_x^i \stackrel{\text{def.}}{=} \frac{\sum_r (\gamma_r + \mu_r) x_{i,r}}{\sum_r [\lambda_r q_r(i) + \gamma_r \sum_j x_{j,r} p^{(r)}(j,i)]}.
$$

By convergence in the Skorohod topology, one means convergence in distribution for the Skorohod topology on the space of trajectories.

PROOF OF THEOREM 1. Recall that if $\mathcal{N}_{\xi_1}$ and $\mathcal{N}_{\xi_2}$, are two independent Poisson processes, and if $M_p(t) = \mathcal{N}_{\xi_p}((0,t]) - \xi_p t$, $p = 1, 2$, are their associated martingales, the latter are orthogonal in the sense that $(M_1(t)M_2(t))$ is a martingale, that is, the bracket process $\langle M_1, M_2 \rangle(t)$ is 0 for all $t \geq 0$; see [10]. The same property holds for stochastic integrals of previsible processes $(H_1(t))$ and $(H_2(t))$—for $t \geq 0$,

$$\left\langle \int_0^\cdot H_1(s)\,dM_1(s), \int_0^\cdot H_2(s)\,dM_2(s) \right\rangle(t) = 0.$$

The increasing process of the renormalized martingale defined above is

$$\langle \overline{M}_{i,r}^N, \overline{M}_{i,r}^N \rangle(t) = \frac{1}{N^2} \langle M_{i,r}^N, M_{i,r}^N \rangle(t),$$

and the increasing process in the right-hand side of the last equation can be evaluated by using the orthogonality of independent Poisson processes mentioned above. By using the fact that, for $(i, r) \in I \times R$ and $t \geq 0$, $X_{i,r}^N(t) \leq \lfloor c_i N \rfloor$, one obtains that there exists some constant $K$ such that

$$\mathbb{E}([M_{i,r}^N(t)]^2) = E(\langle M_{i,r}^N, M_{i,r}^N \rangle(t)) \leq KNt.$$

Doob's inequality implies that the martingale $(\overline{M}_{i,r}^N(t))$ converges a.s. to 0 uniformly on compact sets. Hence the stochastic fluctuations represented by the martingales vanish in limit.

Now, by using the results of Kurtz [9], similarly to their use in Hunt and Kurtz [6] for loss networks, one can prove that any weak limit $X = (X_{i,r})$ of the process $\overline{X}^N$ satisfies the following equations: For $(i,r) \in I \times R$,

$$X_{i,r}(t) = X_{i,r}(0) + \int_0^t \left( \lambda_r q_r(i) + \gamma_r \sum_{j \in I} p^{(r)}(j,i) X_{j,r}(s) \right) \pi_{X(s)}(\overline{\mathbb{N}}_i^I)\,ds$$

(6)

$$- (\gamma_r + \mu_r) \int_0^t X_{i,r}(s)\,ds,$$

where $\overline{\mathbb{N}}_i^I = \{m = (m_j) \in (\mathbb{N} \cup \{+\infty\})^I : m_i \geq 1\}$ and for $x = (x_{ir}) \in \mathcal{X}_c$, $\pi_x$ is *some* stationary probability measure on $\overline{\mathbb{N}}^I = (\mathbb{N} \cup \{+\infty\})^I$ of the Markov jump process whose $Q$-matrix $(B_x(\cdot,\cdot))$ is defined as

$$B_x(m, m - e_i) = \sum_r \lambda_r q_r(i) \quad \text{if } m_i \geq 1,$$

$$B_x(m, m + e_i) = \sum_r x_{i,r} \left( \mu_r + \gamma_r \left( p^{(r)}(i,0) + \sum_{j \in I} p^{(r)}(i,j) \mathbb{1}_{\{m_j = 0\}} \right) \right),$$

$$B_x(m, m - e_i + e_j) = \sum_r \gamma_r x_{j,r} p^{(r)}(j,i) \quad \text{if } m_i \geq 1,$$

where $e_i$ denotes the $i$th unit vector of $\mathbb{R}^I$. Moreover, the probability distribution $\pi_x$ has to satisfy the following condition:

(7) $$\pi_x(m \in \overline{\mathbb{N}}^I : m_i = +\infty) = 1 \qquad \text{if } \sum_r x_{i,r} < c_i.$$

The Markov process $(m^x(t))$ associated with the matrix $B_x(\cdot,\cdot)$ describes the evolution of $Y^N(t/N) = (Y_i^N(t/N))$, that is, the time-rescaled process of the numbers of free units of capacity at different nodes during a time interval $[t, t+Ndt[$ when the renormalized process $\overline{X}^N$ is around $x$ on the normal time scale. Compared to $(X^N(t))$, the process $(Y^N(t))$ indeed evolves on a rapid time scale, so that

$$\int_t^{t+dt} \mathbb{1}_{\{Y_i^N(s^-)>0\}} \, ds \sim \pi_x(\overline{\mathbb{N}}_i^I) \, dt,$$

that is, such quantities can be replaced, in limit, by the average values of indicator functions under some limiting regime $\pi_x$ of $Y^N$ when $\overline{X}^N(t) \sim x$. Hunt and Kurtz [6] provide a detailed treatment of these interesting questions; see also [3, 4] and [12] for the analysis of several examples.

In our case, the marginals of $(m^x(t))$ are also Markov, due to the fact that each customer occupies only one node at a time so that acceptance at node $i$ only depends on the number of free units there. For $i \in I$, the process $(m_i^x(t))$ of the number of free units at node $i$ when the renormalized process is around $x$ is a classical birth and death process on $\overline{\mathbb{N}}$ whose rates are given by

$$q(m, m+1) = N \sum_r (\gamma_r + \mu_r) x_{i,r},$$

$$q(m, m-1) = N \sum_r \left( \lambda_r q_r(i) + \gamma_r \sum_j x_{j,r} p^{(r)}(j, i) \right) \qquad \text{if } m \geq 1.$$

The point $+\infty$ is an absorbing point. Under the condition

(8) $$\sum_r (\gamma_r + \mu_r) x_{i,r} < \sum_r \left( \lambda_r q_r(i) + \gamma_r \sum_j x_{j,r} p^{(r)}(j, i) \right),$$

the geometric distribution with parameter

$$\sum_r (\gamma_r + \mu_r) x_{i,r} \bigg/ \sum_r \left( \lambda_r q_r(i) + \gamma_r \sum_j x_{j,r} p^{(r)}(j, i) \right) = \rho_x^i$$

and $\delta_{+\infty}$, the Dirac distribution at $+\infty$, are the two extreme invariant measures of the process $(m^x(t))$. If $\sum_r x_{i,r} = c_i$ and condition (8) holds, then the quantity $\pi_x(\overline{\mathbb{N}}_i^I)$ is necessarily some convex combination of 1 and $\rho_x^i$. For such an $i \in I$, by summing equations (6) over $r$, it is easy to check that the quantity $\pi_x(\overline{\mathbb{N}}_i^I)$ cannot be more than $\rho_x^i$ (otherwise the finite capacity condition $\sum_r x_{i,r} \leq c_i$ would be violated). One gets that $\pi_x(\overline{\mathbb{N}}_i^I) = \rho_x^i$.

The other cases follow from condition (7) or the transience of the process $(m_i^x(t))$. Since the differential equation (5) clearly has a unique solution, the theorem is proved. $\square$

REMARK. For $t > 0$, the above proof shows that the quantity $\tau_i(x(t))$ can be interpreted as the probability that a call is accepted at node $i$ at time $t$. If the limiting dynamical system has a unique equilibrium point $x$ (which will be shown in the sequel), then by using arguments similar to these in [6], $\tau_i(x)$ can be seen as the limiting stationary probability that a call is accepted at node $i$.

**4. Equilibrium points.** Theorem 1 shows that equilibrium points $x \in \mathcal{X}_c$ of the limiting dynamical system, that is, those $x$ that satisfy $x'_{i,r}(t) = 0$ for any $(i,r) \in I \times R$ and $t \geq 0$ when $(x_{i,r}(0)) = x$, are the solutions of the following set of equations:

$$(9) \quad (\gamma_r + \mu_r)x_{i,r} = \left(\lambda_r q_r(i) + \gamma_r \sum_j x_{j,r} p^{(r)}(j,i)\right)\tau_i(x), \qquad (i,r) \in I \times R,$$

where $\tau_i(x)$ is defined as in Theorem 1. Note that $\tau_i(x) \in (0,1]$ and that either $\tau_i(x) = 1$ or $\sum_r x_{i,r} = c_i$.

4.1. *Characterizations and existence of equilibrium points.* If $x \in \mathcal{X}_c$ satisfies (9), it is a solution of the equations

$$(10) \quad (\gamma_r + \mu_r)x_{i,r} = \left(\lambda_r q_r(i) + \gamma_r \sum_j x_{j,r} p^{(r)}(j,i)\right)t_i \qquad \forall (i,r) \in I \times R,$$

for some $t = (t_i) \in (0,1]^I$ such that for any $i \in I$, either $t_i = 1$ or $\sum_r x_{i,r} = c_i$.

Conversely, if $x \in \mathcal{X}_c$ is a solution of (10) for a fixed $i \in I$, then there are two cases:

- If $\lambda_r q_r(i) + \gamma_r \sum_j x_{j,r} p^{(r)}(j,i) = 0$ for all $r \in R$, then $x_{i,r} = 0$ for all $r$ and thus, necessarily, $\tau_i(x) = 1$ and equations (9) hold trivially.
- Otherwise, by summing these relations over $r \in R$, one gets the identity

$$t_i = \frac{\sum_r (\gamma_r + \mu_r)x_{i,r}}{\sum_r (\lambda_r q_r(i) + \gamma_r \sum_j x_{j,r} p^{(r)}(j,i))}.$$

If $t_i = 1$, then $\rho_x^i = 1$ and so, by definition of $\tau_i(x)$, $\tau_i(x) = 1 = t_i$. If $t_i < 1$, then due to the above assumption, we necessarily have $\sum_r x_{i,r} = c_i$, so

$$\tau_i(x) = \frac{\sum_r (\gamma_r + \mu_r)x_{i,r}}{\sum_r (\lambda_r q_r(i) + \gamma_r \sum_j x_{j,r} p^{(r)}(j,i))} \wedge 1 = t_i.$$

Equations (9) are thus satisfied for $x$.

The following characterization of equilibrium points of the system has thus been obtained:

PROPOSITION 1 (Characterization of equilibrium points). *The equilibrium points of the limiting dynamical system are the elements $x \in \mathcal{X}_c$ such that there exists some $t \in (0,1]^I$ satisfying:*

1. *For any $(i,r) \in I \times R$,*

$$\text{(11)} \qquad x_{i,r} = \left( \alpha_r q_r(i) + \beta_r \sum_j x_{j,r} p^{(r)}(j,i) \right) t_i.$$

2. *For any $i \in I$, either $t_i = 1$ or $\sum_r x_{i,r} = c_i$,*

*where $\alpha_r = \lambda_r/(\gamma_r + \mu_r)$ and $\beta_r = \gamma_r/(\gamma_r + \mu_r)$ for $r \in R$.*

To prove the existence of a fixed point, a second characterization of equilibrium points will be useful:

PROPOSITION 2 (Existence of equilibrium points). *The equilibrium points of the dynamical system* (5) *of Theorem* 1 *are the fixed points in $\mathcal{X}_c$ of the function $\Phi_c$ defined by, for $x \in \mathcal{X}_c$,*

$$\text{(12)} \quad \Phi_c(x) = \left( \Theta_{c_i} \left( \left( \alpha_r q_r(i) + \beta_r \sum_j x_{j,r} p^{(r)}(j,i), r \in R \right) \right), i \in I \right),$$

*where, for $z > 0$ and $u \in [0,+\infty)^R$,*

$$\Theta_z(u) = \left( \frac{z}{\sum_r u_r} \wedge 1 \right) u.$$

*The function $\Phi_c$ has at least one fixed point.*

PROOF. Note that the function $\Theta_c$ maps $[0,+\infty)^R$ into the subset $\{u \in [0,+\infty)^R : \sum_r u_r \leq c\}$ and $\Phi_c(x)$ indeed belongs to $\mathcal{X}_c$: its $(i,r)$th coordinate is 0 whenever $i \notin I_r$.

The characterization of equilibrium points follows from Proposition 1 and by noting that, for $u \in [0,+\infty)^R$, $z > 0$ and $v \in [0,+\infty)^R$ such that $\sum_r v_r \leq z$, there is an equivalence between the identity $\Theta_z(u) = v$ and the fact that there exists some $t \in (0,1]$ such that $v = tu$ and either $t = 1$ or $\sum_r v_r = z$.

The existence of a fixed point is then a consequence of Brouwer's fixed point theorem, since $\mathcal{X}_c$ is a convex compact subset of $\mathbb{R}^{I \times R}$ and $\Phi_c$ is a continuous function from $\mathcal{X}_c$ into itself. □

4.2. *The example of deterministic routes.* Requests of class $r$ use a deterministic route of length $L \in \mathbb{N} \cup \{+\infty\}$ consisting of a sequence $I_r = (i_p, 0 \leq p < L)$ with values in $I$ such that

$$q_r(i_0) = 1, \qquad p^{(r)}(i_p, i_{p+1}) = 1 \qquad \text{for } 0 \leq p < L-1$$

and $p^{(r)}(i_{L-1}, 0) = 1$ if $L < +\infty$. Note that, since $I$ is finite, the case $L = +\infty$ necessarily corresponds to a route $r$ which eventually becomes periodic. Equilibrium points as described in Proposition 1 can be written more explicitly in terms of $t$ solving (11)

$$x_{i,r} = \left(\alpha_r q_r(i) + \beta_r \sum_j x_{j,r} p^{(r)}(j,i)\right) t_i$$

as follows:

1. For a nonperiodic deterministic route, $L < +\infty$, these equations reduce to a recursion—for $0 \leq p < L$,

$$x_{i_p, r} = \alpha_r \beta_r^p \prod_{k=0}^{p} t_{i_k}.$$

2. For a periodic route $r$ consisting of nodes $i_0, i_1, \ldots, i_{k-1}$ and then the infinite loop $i_k, i_{k+1}, \ldots, i_{k+l-1}, i_k, i_{k+1}, \ldots$. Provided that the $(t_{i_k})$ are such that $\beta_r^l t_{i_k} \ldots t_{i_{k+l-1}} < 1$, the solutions are given by

$$x_{i_h, r} = \alpha_r \beta_r^h \prod_{0 \leq m \leq h} t_{i_m}, \qquad 0 \leq h \leq k-1,$$

(13)

$$x_{i_h, r} = \frac{\alpha_r \beta_r^h t_{i_0} t_{i_1} \ldots t_{i_h}}{1 - \beta_r^l t_{i_k} \ldots t_{i_{k+l-1}}}, \qquad h \geq k.$$

The above calculations show that an equilibrium point $(x_{i,r})$ has a polynomial expression in $t = (t_j)$ whose degree is related to the rank of $i$ along the route in the case of a nonperiodic route, and that $(x_{i,r})$ is given by a power series in $t$ when the route $r$ is periodic. Moreover, these quantities have to satisfy the following constraints: for $i \in I$, then either $t_i = 1$ or $\sum_r x_{i,r} = c_i$. The exact expression of fixed points in the case of deterministic routes is therefore very likely to be nontractable. As will be seen, even the uniqueness is not a simple problem.

The complexity of exact expressions is illustrated by a simple example of a network with two nodes, $I = \{1, 2\}$, and two deterministic nonperiodic routes: the first class enters at node 1, goes to node 2 then exits, whereas the second class does the opposite. Take $\mu_1 = \mu_2 = 0$ so that $\beta_1 = \beta_2 = 1$. It is then easy to show that:

1. An equilibrium point associated to $(t_1, t_2)$ with $t_1 = t_2 = 1$ exists if and only if $\alpha_1 + \alpha_2 \leq c_1$ and $\alpha_1 + \alpha_2 \leq c_2$. In this case, $x_{1,1} = x_{2,1} = \alpha_1$ and $x_{1,2} = x_{2,2} = \alpha_2$.
2. An equilibrium point exists with $t_1 = 1$ and $t_2 < 1$ if and only if
$$\alpha_1 + \frac{\alpha_2}{\alpha_1 + \alpha_2} c_2 \leq c_1 \quad \text{and} \quad \alpha_1 + \alpha_2 > c_2.$$

Under these conditions it is then unique:
$$x_{1,1} = \alpha_1, \qquad x_{2,1} = \alpha_1 \frac{c_2}{\alpha_1 + \alpha_2},$$
$$x_{1,2} = x_{2,2} = \alpha_2 \frac{c_2}{\alpha_1 + \alpha_2}.$$

3. By symmetry, analogous results hold with $t_1 = 1$ and $t_2 < 1$.
4. An equilibrium point exists with $t_1 < 1$ and $t_2 < 1$ if and only if
$$\alpha_1 + \frac{\alpha_2}{\alpha_1 + \alpha_2} c_2 > c_1 \quad \text{and} \quad \alpha_2 + \frac{\alpha_1}{\alpha_1 + \alpha_2} c_1 > c_2.$$

In this case the solution is unique:
$$x_{1,1} = \alpha_1 t_1, \qquad x_{2,1} = \alpha_1 t_1 t_2,$$
$$x_{1,2} = \alpha_2 t_1 t_2, \qquad x_{2,2} = \alpha_2 t_2,$$

with
$$t_1 = \frac{(\alpha_1 c_1 - \alpha_2 c_2 - \alpha_1 \alpha_2) + \sqrt{(\alpha_1 c_1 - \alpha_2 c_2 - \alpha_1 \alpha_2)^2 + 4 c_1 \alpha_2 \alpha_1^2}}{2\alpha_1^2},$$

$t_2$ having a similar expression with the subscripts 1 and 2 exchanged.

It is not difficult to check that these four cases are disjoint and cover all situations. Therefore, the uniqueness of the equilibrium point holds in this case.

A similar approach does not seem possible for a more complicated system of deterministic routes. Even proving uniqueness in such a context is challenging.

4.3. *Uniqueness of equilibrium points.* In view of Proposition 2, to prove the uniqueness of equilibrium points, a contraction property of $\Phi_c$ would suffice. But it can be shown that $\Phi_c$ is generally not a contraction for classical norms.

For example, in the simple network considered above with $\beta_1 = \beta_2 = 1$, the equation $\Phi_c(x) = y$ is
$$(y_{1,1}, y_{1,2}) = \Theta_{c_1}(\alpha_1, x_{2,2}) \quad \text{and} \quad (y_{2,1}, y_{2,2}) = \Theta_{c_2}(x_{1,1}, \alpha_2).$$

When $c_1 > \alpha_1$ and $c_2 > \alpha_2$, one can choose $x$ and $x'$ in $\mathcal{X}_c$ such that

$$\begin{cases} \alpha_1 + x_{2,2} \leq c_1, & \alpha_1 + x'_{2,2} \leq c_1, & x_{1,1} + \alpha_2 \leq c_2, \\ x'_{1,1} + \alpha_2 \leq c_2, & x_{1,2} = x'_{1,2}, & x_{2,1} = x'_{2,1}. \end{cases}$$

Then, in this case, $\|\Phi_c(x) - \Phi_c(x')\|_p = \|x - x'\|_p$ for $p \in [1, +\infty]$, where $\|x\|_p$ is the $L_p$-norm $(\|x\|_p)^p = \sum_{i,r} |x_{i,r}|^p$ for $p < +\infty$ and $\|x\|_\infty = \max\{|x_{i,r}| : (i,r) \in I \times R\}$.

Under the condition $\max\{\beta_r : r \in R\} < 1$ and in the case of deterministic nonperiodic routes, the function

$$x \to \left( \alpha_r q_r(i) + \beta_r \sum_j x_{j,r} p^{(r)}(j,i), (i,r) \in I \times R \right)$$

is a contraction for any $L_p$-norm. However, the same property does not necessarily hold for $\Phi_c$, since it can be shown that the function $\Theta_c$, $c > 0$, is not a contraction for any $L_p$-norm on $[0, +\infty)^R$, except when $|R| = 1$, or when $|R| = 2$ and $p = +\infty$.

*A dual approach.* To prove uniqueness in the general case, the point of view is changed—instead of looking for $x \in \mathcal{X}_c$ which are equilibrium points of the limiting dynamics associated to a given vector $c = (c_i, i \in I) \in (0, +\infty)^I$ of capacities, an element $x$ is given and one looks for the set of vectors $c$ such that $x$ is a equilibrium point of the limiting dynamics. The uniqueness of the equilibrium point for a given $c$ is then equivalent to the property that those sets of vectors associated to two different values of $x$ do not intersect.

Define

$$\mathcal{X}_\infty \stackrel{\text{def.}}{=} \{x \in [0, +\infty)^{I \times R} : x_{i,r} = 0 \text{ if } i \notin I_r\}.$$

It is of course enough to consider the solutions $x$ in $\mathcal{X}_\infty$ that satisfy (11) for some $t \in (0,1]^I$. The first step of this analysis is to show that for any $t \in (0,1]^I$, a solution $x$ to the system of equations (11) is at most unique.

PROPOSITION 3 (Probabilistic representation). *If $t \in (0,1]^I$ is such that the system of equations* (11) *has a solution in $\mathcal{X}_\infty$, this solution is unique and can be expressed as*

$$(14) \qquad x_{i,r}^t = \alpha_r \mathbb{E}\left( \sum_{k=0}^{+\infty} \beta_r^k \prod_{p=0}^k t_{Z_p^{(r)}} \mathbb{1}_{\{Z_k^{(r)} = i\}} \right) \qquad \forall (i,r) \in I \times R,$$

*where $(Z_n^{(r)})$ is a (possibly killed) Markov chain with transition matrix $p^{(r)}(\cdot, \cdot)$ and initial distribution $q_r$.*

Note that the above expression for $(x_{i,r})$ generalizes the formula obtained for periodic deterministic Markovian routes since, using the same notation as in the example of periodic deterministic routes, equation (14) gives, for $h \geq k$,

$$x_{i_h} = \alpha_r \beta_r^h t_{i_0} t_{i_1} \cdots t_{i_h} \sum_{j=0}^{+\infty} (\beta_r^p t_{i_k} \cdots t_{i_{k+l-1}})^j = \frac{\alpha_r \beta_r^h t_{i_0} t_{i_1} \cdots t_{i_h}}{1 - \beta_r^p t_{i_k} \cdots t_{i_{k+l-1}}},$$

which is formula (13), and $x_{i_h} = \alpha_r \beta_r^h t_{i_0} \cdots t_{i_h}$ for $h < k$.

PROOF OF PROPOSITION 3. The system of equations (11) splits into $|R|$ subsystems of equations, one for each $r \in R$, with unknown variables $(x_{i,r}, i \in I_r)$. Consider just one of these $|R|$ systems and remove the index $r$ for simplicity. $J$ is defined as the range in $I$ of the Markov chain $(Z_k)$ with initial distribution $q$ and transition matrix $p(\cdot, \cdot)$. Such a subsystem of equations can be expressed as

$$x_i = \left( \alpha q(i) + \beta \sum_j x_j p(j, i) \right) t_i, \qquad i \in J.$$

This system of equations has a solution since the system of equations (11) is assumed to have one. For $i \in J$, set $y_i = x_i/(\alpha t_i)$ (remember that both $\alpha$ and $t_i$ are positive). The vector $y = (y_i)$ then solves the equations

(15) $$y_i = q(i) + \sum_j y_j \widetilde{P}(j, i), \qquad i \in J,$$

with $\widetilde{P}(j, i) = \beta t_j p(j, i)$. The matrix $\widetilde{P} = (\widetilde{P}(i, j))$ is sub-Markovian and $(\widetilde{Z}_n)$ denotes the Markov chain with initial distribution $(q(i))$ and transition matrix $\widetilde{P}$. For $i \in J$, clearly $y_i \geq q(i) = \mathbb{P}(\widetilde{Z}_0 = i)$, and by induction, the above equation gives that, for $n \geq 1$,

$$y_i \geq \mathbb{E}(\mathbb{1}_{\{\widetilde{Z}_0 = i\}} + \mathbb{1}_{\{\widetilde{Z}_1 = i\}} + \cdots + \mathbb{1}_{\{\widetilde{Z}_n = i\}}).$$

By letting $n$ go to infinity, we get

$$y_i \geq u_i \stackrel{\text{def.}}{=} \mathbb{E}\left( \sum_{k=0}^{+\infty} \mathbb{1}_{\{\widetilde{Z}_k = i\}} \right) \qquad \forall i \in J.$$

For any $i \in J$, the above inequality implies that

$$\sum_{k=0}^{+\infty} \widetilde{P}^k(i, i) < +\infty,$$

leading to the conclusion that the state $i$ is transient for the Markov chain $(\widetilde{Z}_n)$.

It is easy to check that $(u_i)$ is also a solution of (15). Consequently, the nonnegative vector $(v_i) = (y_i - u_i)$ satisfies the equation

$$v_i = \sum_j v_j \widetilde{P}(j, i), \qquad i \in J,$$

which is the invariant measure equation for this Markov chain. Since all the states are transient, we necessarily have $v_i = 0$ for all $i \in J$. The uniqueness is thus proved. It is easy to check that the representation of $(x_i)$ in terms of the Markov chain $(Z_n)$ is indeed given by the representation of $(u_i)$ in terms of the Markov chain $(\widetilde{Z}_n)$. The proposition is thus proved. □

DEFINITION 1. The set $\mathcal{T}$ is the subset of $t \in (0,1]^I$ such that the system of equations (11) has a solution, denoted by $x^t = (x_{i,r}^t)$ (it is unique by the above proposition). For $t \in \mathcal{T}$ and $i \in I$, define

$$\sigma_i(t) = \sum_r x_{i,r}^t = \sum_r \alpha_r \mathbb{E}\left(\sum_{k=0}^{+\infty} \beta_r^k \prod_{p=0}^{k} t_{Z_p^{(r)}} \mathbb{1}_{\{Z_k^{(r)} = i\}}\right),$$

where $(Z_n^{(r)})$ is, as before, a Markov chain with transition matrix $p^{(r)}(\cdot, \cdot)$ and initial distribution $q_r$.

LEMMA 1 (Strong monotonicity). *If $t = (t_i)$ and $t' = (t_i')$ are elements of $\mathcal{T}$ such that, for any $i \in I$,*

$$t_i < t_i' \implies \sigma_i(t) \geq \sigma_i(t') \quad \text{and} \quad t_i' < t_i \implies \sigma_i(t') \geq \sigma_i(t),$$

*then $t = t'$.*

PROOF. The assumption on $t$ and $t'$ gives

(16) $$\sum_{i \in I} \log(t_i'/t_i)(\sigma_i(t') - \sigma_i(t)) \leq 0.$$

The definition of $\sigma_i$ gives the following representation for the difference $\sigma_i(t') - \sigma_i(t)$:

$$\sigma_i(t') - \sigma_i(t) = \sum_r \alpha_r \mathbb{E}\left[\sum_{k=0}^{\infty} \beta_r^k \left(\prod_{h=0}^{k} t'_{Z_h^{(r)}} - \prod_{h=0}^{k} t_{Z_h^{(r)}}\right) \mathbb{1}_{\{Z_k^{(r)} = i\}}\right].$$

Note that, as in the proof of Proposition 3, the infinite sums within the expectation are integrable, thereby allowing these algebraic operations. By substituting this expression into (16) and exchanging summations first on $i \in I$ and $r \in R$ and then on $i \in I$ and $k \in \mathbb{N}$ (remembering that $I$ and $R$ are finite), one gets

$$\sum_r \alpha_r \mathbb{E}\left[\sum_{k=0}^{\infty} \beta_r^k \log(t'_{Z_k^{(r)}}/t_{Z_k^{(r)}}) \left(\prod_{h=0}^{k} t'_{Z_h^{(r)}} - \prod_{h=0}^{k} t_{Z_h^{(r)}}\right) \mathbb{1}_{\{Z_k^{(r)} \neq 0\}}\right] \leq 0$$

and, by extending the definitions of $t$ and $t'$ to the coordinate 0 so that $t_0 = t'_0 = 1$,

$$\sum_r \frac{\alpha_r}{\beta_r} \mathbb{E}\left[\sum_{k=0}^{\infty} \log(\beta_r t'_{Z_k^{(r)}}/\beta_r t_{Z_k^{(r)}})\left(\prod_{h=0}^{k} \beta_r t'_{Z_h^{(r)}} - \prod_{h=0}^{k} \beta_r t_{Z_h^{(r)}}\right)\right] \leq 0.$$

Proposition A.1 of the Appendix applied to the expression inside the expectation implies that, with probability 1, this integrand should be 0. Consequently, the same proposition implies that for any $r \in R$, the identity $t_{Z_k^{(r)}} = t'_{Z_k^{(r)}}$ holds almost surely for any $k \in \mathbb{N}$. Hence, $t_i = t'_i$ for any $i \in I_r$ and any $r \in R$ by definition of $I_r$. One concludes that $t = t'$, since $I = \bigcup_r I_r$. The lemma is thus proved. □

The main result concerning the equilibrium points of the limiting dynamical system (5) can now be established:

THEOREM 2 (Uniqueness of equilibrium points). *There is a unique equilibrium point of the dynamical system $(x_{i,r}(t), (i,r) \in I \times R)$ defined by (5).*

PROOF. For $t \in \mathcal{T}$, define $C_t$ as the set of vectors $c = (c_i) \in (0, +\infty[^I$ such that $x^t$ is a fixed point of the dynamical system associated with capacities $(c_i)$. For $t \in \mathcal{T}$ and $c \in (0, +\infty)^I$, Proposition 1 shows that if $c \in C_t$ then, for any $i \in I$, $\sigma_i(t) \leq c_i$, and when $t_i < 1$ then $\sigma_i(t) = c_i$.

For $t, t' \in \mathcal{T}$, assume that there exists some $c \in C_t \cap C_{t'}$. If $i \in I$, the relation $t_i < t'_i$ implies that $t_i < 1$ and therefore that $\sigma_i(t') \leq \sigma_i(t) = c_i$. From Lemma 1, one concludes that, necessarily, $t = t'$. The uniqueness of equilibrium points readily follows from the result that if $z$ and $z'$ are equilibrium points of the dynamical system (5) associated with some vector of capacities $c \in (0, +\infty)^I$, then there exist $t$ and $t' \in \mathcal{T}$ such that $z = x^t$ and $z' = x^{t'}$. Since $c \in C_t \cap C_{t'}$, we have $t = t'$ and therefore $z = z'$. The theorem is thus proved. □

## APPENDIX

This section is devoted to the proof of a key technical result for the proof of the uniqueness of equilibrium points. It involves an expression which bears some similarity to a relative entropy.

PROPOSITION A.1. *Let $u = (u_i)_{i \in \mathbb{N}}$ and $u' = (u'_i)_{i \in \mathbb{N}}$ be two sequences of elements of $(0,1]$. If the series*

$$\sum_{i=0}^{+\infty} \log(u'_i/u_i)\left(\prod_{j \leq i} u'_j - \prod_{j \leq i} u_j\right)$$

*converges, then its sum is nonnegative and equals 0 if and only if $u = u'$.*

PROOF. It is first proved by induction on $n \in \mathbb{N}$ that for any $u, u' \in (0,1]^n$,

$$(A.1) \qquad f_n(u,u') \stackrel{\text{def.}}{=} \sum_{i=0}^n \log(u'_i/u_i) \left( \prod_{j \leq i} u'_j - \prod_{j \leq i} u_j \right) \geq 0.$$

This is obviously true for $n=0$. Now assume this inequality holds for any integer $k < n$. Let $u$ and $u'$ be some fixed elements of $(0,1]^n$.

- If there exists some $k$ such that $1 \leq k \leq n$ and

$$\left( \prod_{j \leq k-1} u'_j - \prod_{j \leq k-1} u_j \right) \left( \prod_{j \leq k} u'_j - \prod_{j \leq k} u_j \right) \leq 0,$$

then $f_n(u,u')$ can be decomposed as follows:

$$f_n(u,u') = f_{k-1}((u_0,\ldots,u_{k-1}),(u'_0,\ldots,u'_{k-1}))$$
$$(A.2) \qquad + f_{n-k}\left( \left( \prod_{j \leq k} u_j, u_{k+1},\ldots,u_n \right), \left( \prod_{j \leq k} u'_j, u'_{k+1},\ldots,u'_n \right) \right)$$
$$- \log\left( \prod_{j \leq k-1} u'_j \bigg/ \prod_{j \leq k-1} u_j \right) \left( \prod_{j \leq k} u'_j - \prod_{j \leq k} u_j \right).$$

From the induction hypothesis and the assumption on $k$, all terms of the right-hand side of this identity are nonnegative, so $f_n(u,u') \geq 0$.

- Otherwise, for any $0 \leq k \leq n$, the quantity $\prod_{j \leq k} u'_j - \prod_{j \leq k} u_j$ has a constant sign and is not 0 (positive, say). There are two cases:

1. If $u_k \leq u'_k$ for all $k$ such that $0 \leq k \leq n$, all terms in the sum defining $f_n(u,u')$ are nonnegative, and hence $f_n(u,u') \geq 0$.
2. If not, let $k \leq n$ be the first index such that $u_k > u'_k$. Since $u_0 < u'_0$, we have $k \geq 1$ and can write

$$f_n(u,u') = f_{n-1}[(u_0,\ldots,u_{k-2},u_{k-1}u_k,u_{k+1},\ldots,u_n),$$
$$(u'_0,\ldots,u'_{k-2},u'_{k-1}u'_k,u'_{k+1},\ldots,u'_n)]$$
$$+ \log(u'_{k-1}/u_{k-1}) \left( (1-u'_k) \prod_{j \leq k-1} u'_j - (1-u_k) \prod_{j \leq k-1} u_j \right).$$

The first term is nonnegative from the induction hypothesis. The second one is also nonnegative, since $u_{k-1} \leq u'_{k-1}$, $u'_k \leq u_k$ and $\prod_{j \leq k-1} u_j \leq \prod_{j \leq k-1} u'_j$. Therefore, $f_n(u,u') \geq 0$ also holds in this case. The proof by induction is thus completed.

Inequality (A.1) is thus true for any $n \in \mathbb{N}$, implying that for any $u, u' \in (0,1]^{\mathbb{N}}$,

$$f_\infty(u,u') \stackrel{\text{def.}}{=} \sum_{i=0}^{+\infty} \log(u_i'/u_i)\left(\prod_{j\leq i} u_j' - \prod_{j\leq i} u_j\right) \geq 0$$

whenever the series converges. The first part of the proposition is thus proved.

Assume now that $f_\infty(u,u') = 0$ for some $u, u' \in (0,1]^{\mathbb{N}}$ such that the series converges. Using the same kind of decomposition as in equation (A.2), $f_\infty(u,u')$ can be expressed as, for some fixed $k \geq 1$,

$$f_\infty(u,u') = f_{k-1}((u_0,\ldots,u_{k-1}),(u_0',\ldots,u_{k-1}'))$$
$$+ f_\infty\left(\left(\prod_{j\leq k} u_j, u_{k+1}, \ldots\right), \left(\prod_{j\leq k} u_j', u_{k+1}', \ldots\right)\right)$$
$$- \log\left(\prod_{j\leq k-1} u_j' \Big/ \prod_{j\leq k-1} u_j\right)\left(\prod_{j\leq k} u_j' - \prod_{j\leq k} u_j\right) = 0.$$

The second term of the right-hand side is clearly well defined, since $f_\infty(u,u')$ is. The first and second terms being nonnegative, we have

$$\log\left(\prod_{j\leq k-1} u_j' \Big/ \prod_{j\leq k-1} u_j\right)\left(\prod_{j\leq k} u_j' - \prod_{j\leq k} u_j\right) \geq 0.$$

Consequently, the difference $u_0' u_1' \cdots u_k' - u_0 u_1 \cdots u_k$ has a constant sign for any $k \in \mathbb{N}$. It can be assumed that these expressions are nonnegative.

1. If $u_i \leq u_i'$ holds for any $i \geq 0$, then each term of the infinite sum defining $f_\infty(u,u')$ is nonnegative and therefore null, since $f_\infty(u,u') = 0$. It clearly implies that $u_i = u_i'$ for all $i \in \mathbb{N}$.
2. Otherwise, since $u_0 \leq u_0'$, define $n \geq 1$ as the smallest integer such that $u_n > u_n'$. Since $u_0 u_1 \cdots u_n \leq u_0' u_1' \cdots u_n'$, there exists some index $i < n$ satisfying $u_i < u_i'$. Define $k$ as the largest such index. In particular, for $k < i < n$, one has $u_i = u_i'$. Therefore,

$$f_\infty(u,u') = f_\infty\left(\left(u_0,\ldots,u_{k-1}, \prod_{j=k}^n u_j, u_{n+1}, \ldots\right),\right.$$
$$\left.\left(u_0',\ldots,u_{k-1}', \prod_{j=k}^n u_j', u_{n+1}', \ldots\right)\right)$$
$$+ \log(u_k'/u_k)\left(\left(1 - \prod_{k<j\leq n} u_j'\right)\prod_{j\leq k} u_j' - \left(1 - \prod_{k<j\leq n} u_j\right)\prod_{j\leq k} u_j\right)$$
$$= 0.$$

The first term is nonnegative and it is easily checked by using the definitions of $k$ and $n$ that the second one is positive. This equality is therefore absurd. This second case is not possible.

The proposition is thus proved. □

N. ANTUNES
UNIVERSIDADE DO ALGARVE
FACULDADE DE CIÊNCIAS E TECNOLOGIA
CAMPUS DE GAMBELAS
8005-139 FARO
PORTUGAL
E-MAIL: nantunes@ualg.pt

C. FRICKER
P. ROBERT
INRIA
DOMAINE DE VOLUCEAU
B.P. 105
78153 LE CHESNAY CEDEX
FRANCE
E-MAIL: Christine.Fricker@inria.fr
Philippe.Robert@inria.fr

D. TIBI
UNIVERSITÉ PARIS 7
UMR 7599
2 PLACE JUSSIEU
75251 PARIS CEDEX 05
FRANCE
E-MAIL: Danielle.Tibi@math.jussieu.fr